\documentclass[12pt]{article}

\usepackage{amssymb}

\usepackage{verbatim}

\def\hH {{\widehat H}_h}

\def\bR {{\mathbb{R}}}
\def\bN {{\mathbb{N}}}

\def\cA {{\mathcal A}}

\def\cH {{\mathcal H}}

\def\cV {{\mathcal V}}

\def\cO {{\mathcal O}}

\def\Di {\displaystyle}
\def\hbr {\hslash}
\def\ccH {{\widehat H}^\hslash}
\def\tH {{\widetilde H}^\hslash}

\def\MR {\mathrm }
\def\tG {{\widetilde g}}

\makeatletter
\@addtoreset{equation}{section}

\makeatother

\newtheorem{theorem}{Theorem}[section]
\newtheorem{lemma}[theorem]{Lemma}
\newtheorem{proposition}[theorem]{Proposition}

\newtheorem{remark}[theorem]{Remark}

\begin{document}

\bibliographystyle{plain}

\begin{center}
\Large \bf {
 Accuracy on  eigenvalues \\
 for a  Schr\"odinger operator
 with a degenerate potential\\
in the semi-classical limit }
\end{center}

\vskip 0.5cm

 \centerline {\bf { Abderemane MORAME$^{1}$
 and  Fran{\c c}oise TRUC$^{2}$}}

{\it {$^{1}$ Universit\'e de Nantes,
Facult\'e des Sciences,  Dpt. Math\'ematiques, \\
UMR 6629 du CNRS, B.P. 99208, 44322 Nantes Cedex 3, (FRANCE), \\
E.Mail: morame@math.univ-nantes.fr}}

{\it {$^{2}$ Universit\'e de Grenoble I, Institut Fourier,\\
            UMR 5582 CNRS-UJF,
            B.P. 74,\\
 38402 St Martin d'H\`eres Cedex, (France), \\
E.Mail: Francoise.Truc@ujf-grenoble.fr }}

  \footnote{
  {\sl Mathematical Classification }~: 35P20.}

\begin{abstract}
We consider a semi-classical Schr\"odinger operator
       $-h^2\Delta +V$  with a degenerate potential
V(x,y) =f(x) g(y) .\\
 g is assumed to be a homogeneous positive function of m variables
 and
 f is a strictly positive function of n variables, with a strict
minimum.  We give  sharp asymptotic
behaviour
of  low eigenvalues bounded by some power
of the parameter h, by improving Born-Oppenheimer approximation.
\end{abstract}

\section{Introduction}

In our paper \cite{MoTr}  we have considered the
Schr\"odinger operator on

$L^2(\bR ^n_x\times \bR ^m_y)$
\begin{equation}\label{defOpH}
\hH \; =\; h^2D^2_x\; +\; h^2D^2_y\; +\; f(x)g(y)
\end{equation}
with $\ g\; \in \; C^\infty (\bR ^m\setminus \{ 0\} )\ $
homogeneous of degree $a>0\; ,$
\begin{equation}\label{Hypg}
g(\mu y)\; =\; \mu ^ag(y)\; >\; 0\; ,\quad
\forall \; \mu >0\ {\rm and}\ \forall \; y\; \in \;
\bR ^m\setminus \{ 0\} \; .
\end{equation}

$h>0\ $ is a semiclassical parameter we assume
to be small.

We have investigated the asymptotic behavior
of the number of eigenvalues less then
$\lambda $ of $\hH \; ,$
\begin{equation}\label{defnlambda}
 N(\lambda ,\ \hH )=
tr(\chi_{]-\infty ,\lambda[}(\hH )
\; =\; \sum_{\lambda_k(\hH)<\lambda} 1\ \; .
\end{equation}
$(tr(P)\ $ denotes the trace of the operator $\ P\; )\; .$ \\
If $\ P\ $ is a self-adjoint operator on a
Hilbert space $\cH \; $, we denote respectively by $\ sp(P)\; ,\ sp_{ess}(P)\ $ and $\ sp_d(P)\ $
the spectrum, the essential spectrum and the
discret spectrum of $\ P\;$.

When $\ -\infty
\; <\; \inf \; sp(P)\; <\;  \inf \;
sp_{ess}(P)\;$ , we denote by $\quad (\lambda_k(P))_{k>0}\ $
 the increasing sequence of
eigenvalues of $\ P\; ,$ repeated according
to their multiplicity:
\begin{equation}\label{deflambdakp}
sp_d(P)\; \bigcap \; ]-\infty ,\ \inf\; sp_{ess}(P)[\; =\; \{ \lambda_k(P)\} \;.
\end{equation}

In this paper we are interested in a sharp estimate
for some eigenvalues of $\hH \; .$
 We make the following assumptions on the other multiplicative part
of the potential:

\begin{equation}\label{Hypf}
\begin{array}{ll}
f\; \in \; C^\infty (\bR ^n),
\ \forall  \alpha \in \bN ^n,\ (|f(x)|+1)^{-1}\partial_x^\alpha f(x)\;
\in L^\infty (\bR ^n)\\
 0\; <\; f(0)\; =\; \inf_{x\in \bR ^n} \; f(x)\\
 f(0)\; <\; \liminf_{|x|\to \infty}\; f(x) \; =\; f(\infty)\\
 \partial ^2f(0)\; >\; 0\\
 \end{array}
 \end{equation}
 $\partial ^2 f(a)$ denotes the hessian matrix:
 $$\partial ^2 f(a)\; =\; \left ( \frac{\partial ^2}{\partial x_i \partial x_j}
 f(a)\right )_{1\leq i,j\leq n}\; .$$
 By dividing $\hH \ $ by $f(0)\; ,$ we can change the parameter $\ h\ $ and
assume that
 \begin{equation}\label{fzero}
 f(0)=1\; .
 \end{equation}

Let us define~:
$ \hbr \; =\; h^{2/(2+a)}$ and   change $y$ in $y\hbr$;
 we can use the homogeneity of $g$ (\ref{Hypg}) to get~:
 \begin{equation}\label{defHcc}
 sp\; (\hH )\; =\; \hbr^a\;
 sp\; (\ccH)\; ,
 \end{equation}
 with $\Di \ \ccH \; =\; \hbr ^2D_x^2\; +\; D_y^2\; +\; f(x)g(y)\;
 =\; \hbr ^2D_x^2\; +\;Q(x,y,D_y)\ :$
 $$Q(x,y,D_y)\; =\; D_y^2\; +\; f(x)g(y)\; .$$
Let us denote the increasing sequence of eigenvalues
 of $\ D^2_y\; +\; g(y)\; ,$
 (on $L^2(\bR ^m)\; )\; ,$ by $\ (\mu_j)_{j>0 }\; .$\\
 The associated  eigenfunctions will be denoted
 by $\ (\varphi_j)_j\; :$
 \begin{equation}\label{defmuk}
 \begin{array}{ll}
 D^2_y\varphi_j (y)\; +\; g(y)\varphi_j(y)\;
 =\; \mu_j \varphi_j(y)\\
 \langle \varphi_j\; |\; \varphi_k \rangle \; =\; \delta_{jk}
 \end{array}
 \end{equation}
 and $\ (\varphi_j)_j\ $ is a Hilbert base of
 $L^2(\bR ^m)\; .$

 By homogeneity (\ref{Hypg}) the eigenvalues
 of $Q_x(y,D_y)=\ D^2_y\; +\; f(x)g(y)\; ,$
 on $L^2(\bR ^m)\; )\; $, for a fixed $x$, are given by the sequence $(\lambda_j(x))_{j>0 }$, where~:
 $\lambda_j(x)=\mu_j\ f^{2/(2+a)}(x) \; .$
 \\
 So as in \cite{MoTr} we get~ :

 \begin{equation}\label{estiH1}
 \ccH\; \geq\; \left [ \;
 \hbr^{2}D^2_x\; +\; \mu_1 f^{2/(2+a)}(x)\; \right ] \; .
 \end{equation}
 This estimate is sharp as we will see below.

Then using  the same kind of estimate as (\ref{estiH1}), one can see that
 \begin{equation}\label{infSpess}
 \inf \; sp_{ess}(\ccH )\; \geq \; \mu_1f^{2/(2+a)}(\infty)\; .
 \end{equation}

 We are in the Born-Oppenheimer approximation situation described
 by A. Martinez in \cite{Ma}~: the "effective "
 potential is given by $\lambda_1(x)=\mu_1\ f^{2/(2+a)}(x)$, the first eigenvalue of $Q_x$,
 and the assumptions on $f$ ensure that this potential admits one
 unique and nondegenerate  well  $U=\{ 0\} $,
  with minimal value equal to $\mu_1$. Hence we can apply theorem 4.1 of
\cite{Ma} and get~:

 \begin{theorem}\label{thGround}
 Under the above assumptions, for any arbitrary $C>0$,
 there exists $\ h_0>0 $ such that, if
 $0<\hbr<h_0\; ,$  the operator $(\ccH)$
 admits a finite number of eigenvalues
 $E_{k}(\hbr )$ in $[\mu_1,\mu_1+C\hbr ]$, equal to the number
 of the eigenvalues $e_k$ of
 $\  D_x^2 \; +\;  \frac{\mu_1}{2+a} <~\partial^2f(0)\ x,\ x\ >\ $
   in  $[0,+C]$ such that~:

 \begin{equation}\label{equaGround}
 E_{k}(\hbr )=\lambda_k(\ccH )\; =\; \lambda_k \left ( \hbr^2
 D^2_x+\mu_1f^{2/(2+a)}(x)\right )\; +\; {\bf O}(\hbr^2)
 \; .
 \end{equation}
 More precisely
 $ E_{k}(\hbr )=\ \lambda_k(\ccH)\; $
 has an asymptotic expansion
 \begin{equation}\label{asympHH}
  E_{k}(\hbr )\  \sim \;
 \mu_1\; +\;\hbr\  (\ e_k\ +\  \sum_{j\geq 1}
 \alpha_{kj}\hbr ^{j/2}\;) .
 \end{equation}
 If $ E_{k}(\hbr )\ $ is asymptotically
 non degenerated, then there exists a quasimode
 \begin{equation}\label{quasiHH}
 \phi_k^\hbr (x,y)\; \sim \;
 \hbr ^{-m_k} e^{-\psi (x)/\hbr}
 \sum_{j\geq 0} \hbr^{j/2}a_{kj}(x,y)\; ,
 \end{equation}
 satisfying
 \begin{equation}\label{quasiHHb}
 \begin{array}{ll}
 C^{-1}_{0} \leq \| \hbr^{-m_k}e^{-\psi (x)/\hbr}a_{k0}(x,y)\| \leq C_0\\
 \| \hbr ^{-m_k}e^{-\psi (x)/\hbr}a_{kj}(x,y)\| \leq C_j\\
 \| \left ( \ccH -\mu_1-\;  \hbr e_k - \sum_{1\leq j\leq J}
 \alpha_{kj}\hbr ^{j/2}
 \right ) \\
 \hbr^{-m_k}e^{-\psi (x)/\hbr}
 \sum_{0\leq j\leq J}\hbr ^{j/2}a_{kj}(x,y)\| \; \leq \; C_J \hbr ^{(J+1)/2}
 \end{array}
 \end{equation}
 \end{theorem}
 The formula (\ref{asympHH}) implies
 \begin{equation}\label{asymptGr}
 \lambda_k(\ccH)\; =\; \mu_1\; +\; \hbr \lambda_k
 \left ( D_x^2 + \frac{\mu_1}{2+a}<\ \partial ^2f(0)\ x\ ,\ x\ >\right )
 \; +\; {\bf O}(\hbr ^{3/2})\; ,
 \end{equation}
 and when $k=1\; ,$ one can improve ${\bf O}(\hbr ^{3/2})\; $
 into ${\bf O}(\hbr ^2)\; .$
The function $\psi$ is defined by ~:
$\psi (x)\ =\ d(x, 0)\ $, where $d$ denotes the Agmon distance related to the degenerate
metric $\mu_1\ f^{2/(2+a)}(x) dx^2.$

\section{Lower energies}

We are interested now with the lower
 energies of $\ccH$ .
 Let us make the change of variables
\begin{equation}\label{changeV}
(x,\; y)\; \to \; (x,\; f^{1/(2+a)}(x)y)\ .
\end{equation}
The Jacobian of this diffeomorphism is $ f^{m/(2+a)}(x)$, so we perform the change
 of test functions~:
$\Di \; u\; \to \; f^{-m/(4+2a)}(x)u\; ,$ 
  to get a unitary transformation.

Thus we get that
\begin{equation}\label{sptH}
sp\; (\ccH )\; =\; sp\; (\tH )
\end{equation}
where $\tH \ $ is the self-adjoint operator on
$L^2(\bR ^n\times \bR ^m)\ $ given by
\begin{equation}\label{deftH}
\tH \; =\; \hbr ^2L^\star (x,y,D_x,D_y)
L(x,y,D_x,D_y)
\; +\;
f^{2/(2+a)}(x)\left ( D^2_y+g(y)\right )\; ,
\end{equation}
with
$$
L(x,y,D_x,D_y)\; =\;
D_x\; +\; \frac{1}{(2+a)f(x)}[(yD_y)
\; -\; i\frac{m}{2}]\nabla f(x)\; .
$$
We decompose $\tH $ in four parts~:
\begin{equation}\label{deftH2}
\begin{array}{ll}
\tH \; =\;
\hbr ^2D^2_x\; +\;
f^{2/(2+a)}(x)\left (
D^2_y+g(y)\right )
\\
+\hbr ^2\frac{2}{(2+a)f(x)}(\nabla f(x)D_x)(yD_y)
\\
+i\hbr ^2\frac{1}{(2+a)f^2(x)}\left ( |\nabla f(x)|^2-
f(x)\Delta f(x)\right )[(yD_y)\; -\; i\frac{m}{2}]
\\
+\; \hbr ^2\frac{1}{(2+a)^2f^2(x)}
|\nabla f(x)|^2[(yD_y)^2\; +\; \frac{m^2}{4}]
\end{array}
\end{equation}

Our goal is to prove that the only significant role up to order 2 in $\hbr$ will be played
by the first operator, namely~:
$\tH_1\ =\
\hbr ^2D^2_x\; +\;
f^{2/(2+a)}(x)\left (
D^2_y+g(y)\right )\ $.

 Let us denote by $\nu_{j,k}^\hbr$ the  eigenvalues of the operator
 $\hbr ^2D^2_x\; +\;\mu_j f^{2/(2+a)}(x) \ $
 and by $\psi_{j,k}^\hbr $   the associated normalized
eigenfunctions  .

Let us consider the following test functions~:
$$ u_{j,k}^\hbr(x,y)= \psi_{j,k}^\hbr(x) \varphi_{j}(y)\ , $$
where the $\varphi_{j}$'s are the eigenfunctions defined in (\ref{defmuk});
we have immediately~:
$$\tH_1( u_{j,k}^\hbr(x,y))\ =\nu_{j,k}^\hbr u_{j,k}^\hbr(x,y)\ .$$
 We will need the following lemma~:

\begin{lemma}\label{ppty} .
  For any integer $\ N\; ,$ there exists a positive constant $C$
depending only on $N$ such that for any  $\ k\leq N\; ,$
 the eigenfunction  $\psi_{j,k}^\hbr$ satisfies the following inequalities ~:
 for any $\ \alpha \; \in \; \bN ^n\; ,\ |\alpha |\; \leq \; 2\; ,$
\begin{equation}\label{ineq}
\begin{array}{ccl}
\|\; \hbr _{j}^{|\alpha |/2}\;|D_x^\alpha \;\psi_{j,k}^\hbr| \; \|
&<& \;C\\
\|\; \left ( \frac{\nabla f(x)}{f(x)}\right ) ^\alpha
\;\psi_{j,k}^\hbr \; \|
&<& \;\hbr _{j}^{|\alpha |/2} C\\
\end{array}
\end{equation}
with $\ \hbr_j\; =\; \hbr \mu_{j}^{- 1/2}\; .$
\end{lemma}
{\bf Proof.}

Let us recall that it is well known, (see \cite{HeSj1} ),
 that
$$ \forall\;  k\; \leq \; N \; ,\quad
\mu_{j}^{-1}\nu^{\hbr}_{j,k}\; =\; 1\; +\;
{\bf O}(\hbr_j)\; .$$
\indent
It is clear also that
\begin{equation}\label{preAgEst}
\left [\hbr_{j}^{2}D_x^2\; +\; f^{2/(2+a)}(x)\; -\; \mu_{j}^{-1}\nu^{\hbr}_{j,k}
\right ] \psi^{\hbr}_{j,k}(x)\; =\; 0\; .
\end{equation}
 We shall need the following inequality, that we can derive easily from (\ref{preAgEst}) and
 the Agmon estimate (see \cite{HeSj1})  ~: $\ \forall \;
\varepsilon \; \in \; ]0,1[\; ,$
\begin{equation}\label{AgEst}
\begin{array}{c}
\varepsilon \int \left [f^{2/(2+a)}(x) - \mu_{j}^{-1} \nu_{j,k}^{\hbr}
 \right ]_{+} e^{2(1-\varepsilon )^{1/2}d_{j,k}(x)/\hbr_j}|\psi_{j,k}^{\hbr}(x)|^2\; dx
 \; \leq \\
\int \left [f^{2/(2+a)}(x) - \mu_{j}^{-1}\nu_{j,k}^{\hbr}
 \right ]_{-} |\psi_{j,k}^{\hbr}(x)|^2\; dx\; ,
\end{array}
\end{equation}
where $\ d_{j,k}\ $ is the Agmon distance associated to the metric
  $ \left [ f^{2/(2+a)}(x) -\mu_{j}^{-1} \nu_{j,k}^{\hbr} \right ]_{+}dx^2 \; .$

Let us prove the lemma for   $ |\alpha |\; =\ 1.$

As $\displaystyle \ \int \left [ \hbr _j^2 |D_x \;
\psi_{j,k}^{\hbr}(x)|^2 +(f^{2/(2+a)}(x) - \mu_{j}^{-1}
\nu_{j,k}^{\hbr})|\psi_{j,k}^{\hbr}(x)|^2\right ] dx\; =\; 0
\; ,$\\
 $\ \mu_{j}^{-1}\nu_{j,k}^{\hbr}-1\; =\; {\bf O}(\hbr_j)\; ,$
 and $\ f^{2/(2+a)}(x)-1\; >\; 0\; , $\\
we get that $\ \hbr_j\|  \; |D_x \; \psi_{j,k}^{\hbr}(x)| \;
\| ^2\; \leq \; C\; .$

Furthermore, we use that $\ C^{-1}| \nabla f(x) | ^2\; \leq\;
f^{2/(2+a)}(x) -1 \; \leq C |\nabla f(x)|^2\; ,$\\
 for
$\ |x|\; \leq \; C^{-1}\; ,$  the exponential decreasing (in $\hbr_j)$
of $\ \psi^{\hbr}_{j,k}\ $ given by (\ref{AgEst}) and the boundness
of $\ |\nabla f(x)|/f(x)\ $ to get
$$\| \frac{|\nabla f(x)|}{f(x)}\psi_{j,k}^{\hbr}(x) \| ^2
\; \leq \; C \int [ f^{2/(2+a)}(x)-1]\; |\psi_{j,k}^{\hbr}(x)|^2\; dx
\; \leq \; \hbr_j C\; .$$

Now we study the case  $ |\alpha |\; =\; 2\ .$

 If $\ c_0\in ]0,1]\ $ is large enough and
 $\ |x|\; \in  \; [\hbr _{j}^{1/2}c_0, \ 2c_{0}]\; ,$ then  we have  
\begin{equation}\label{cou}
|x|^2/C\; \leq \; f^{2/(2+a)}(x) - \mu_{j}^{-1}\nu_{j,k}^{\hbr} \; \leq
\; C |x|^2
\end{equation}

Therefore there exists
 $\ C_1>1\ $ such that $  C_{1}^{-1} |x|^2\; \leq \; d_{j,k}(x)\; \leq \; C_1|x|^2\; ,$\\
and then
 \begin{equation}\label{coun}
 |x|^2\; \leq \; \hbr_j Ce^{d_{j,k}(x)/\hbr_j }\; .
\end{equation}

 Then the inequality~: $\ C^{-1}|x| \leq\; |\nabla f(x)|\; \leq \; C|x|\; .$
 together with (\ref{cou}) , (\ref{coun}) and (\ref{AgEst}) entail that
 $$\begin{array}{ll} \int_{|x|\geq C_0\hbr _{j}^{12}} \frac{|\nabla f(x)|^4}{f^4(x)}
 |\psi_{j,k}^{\hbr}(x)|^2\; dx &  \\
  & \leq \; \hbr _jC
 \int \left [f^{2/(2+a)}(x) - \mu_{j}^{-1} \nu_{j,k}^{\hbr}
 \right ]_{+} e^{d_{j,k}(x)/\hbr_j}|\psi_{j,k}^{\hbr}(x)|^2\; dx
 \\
  & \leq \; \hbr _jC
 \int \left [f^{2/(2+a)}(x) - \mu_{j}^{-1} \nu_{j,k}^{\hbr}
 \right ]_{-} |\psi_{j,k}^{\hbr}(x)|^2\; dx
 \\
  & \leq \; \hbr _j^2C\; .
 \end{array} $$

 It remains to estimate $\ \hbr_j^2\| D_x^{\alpha}\psi_{j,k}^{\hbr}(x)\| \; $ with
  $ |\alpha |\; =\; 2\ .$
 \\
 We use that $\ -\hbr _j^2\Delta \psi_{j,k}^{\hbr}(x)\; =\;
 [-f^{2/(2+a)}(x) + \mu_{j}^{-1} \nu_{j,k}^{\hbr}
 ]\psi_{j,k}^{\hbr}(x)\; ,$\\
 and that we have proved that $\ \|
 [-f^{2/(2+a)}(x) + \mu_{j}^{-1} \nu_{j,k}^{\hbr}
 ]\psi_{j,k}^{\hbr}(x)\| \; \leq \; \hbr_j C\; ;$\\
 so $\ \| D_x^\alpha \psi_{j,k}^{\hbr}(x)\| \; \leq \; C/\hbr_j \; $
 if $\ |\alpha |=2\; .$

We will need the following result.

\begin{proposition}\label{thHorm}

Let $\ V(y)\; \in \; C^\infty (\bR ^m)\ $ such that
\begin{equation}\label{eqHorm}
\begin{array}{l}
\exists\;  s\; >\; 0\; ,\ C_0\; >\; 0\ s.t.\quad
-C_0 + |y|^s/C_0\; \leq \; V(y)\; \leq \; C_0(|y|^s+1)\\
\forall \; \alpha \in \bN ^m\; ,\ (1+|y|^2)^{(s-|\alpha |)/2}\partial _y^\alpha
V(y)\; \in \; L^\infty (\bR ^m)\; .
\end{array}
\end{equation}
If
$\ u(y)\; \in \; L^2(\bR ^m)\ $ and
$\ D^2_yu(y)\; +\; V(y)u(y)\; \in \; S(\bR ^m)\; ,$\\
 then $\ u\; \in \; S(\bR ^m)\; .$
$\; (\ S(\bR ^m)\ $ is the Schwartz space).
\end{proposition}

The proof comes from the fact that there exists a parametrix of $\
D_y^2+V(y)\ $ in some class of pseudodifferential operator: see
for the more general case in \cite{Hor}, or for this special case
in Shubin book \cite{Shu}.


\begin{theorem}\label{thGround2} .

Under the assumptions (\ref{Hypg}) and (\ref{Hypf}),
for any fixed integer $\ N\ >0\ ,$
there exists a positive constant $h_0(N)$ verifying~:
for any $ \hbr \in ]0, h_0(N)[$, for any $k\leq N\; $ and any
$j\leq N\; $  such that
 $$\ \mu_j\; <\; \mu_1f^{2/(2+a)}(\infty)\; ,$$
 there exists an eigenvalue
$\; \lambda_{jk}\; \in \; sp_d \; (\ccH )\ $
 such that
 \begin{equation}\label{equaGround2}
 |\; \lambda_{jk}\; -\; \lambda_k \left ( \hbr^2 D^2_x+\mu_jf^{2/(2+a)}(x)\right )\; |\; \leq \; \hbr ^{2}C \; .
 \end{equation}

 Consequently, when $k=1\; ,$ we have
 \begin{equation}\label{equaGround2b}
 |\; \lambda_{j1}\; -\; \left [\mu_j\; +\; \hbr (\mu_j)^{1/2}
 \frac{tr((\partial ^2f(0))^{1/2})}{(2+a)^{1/2}} \right ]\; |\; \leq \;\hbr ^{2}C \; .
 \end{equation}
 \end{theorem}

{\bf Proof ~.}

The first part of the theorem will follow if we  prove that~:
$$
\|(\ccH\ -\ \tH_1)( u_{j,k}^\hbr(x,y))\ \| \; =\;
\| (\ccH\ -\ \nu_{j,k}^\hbr) u_{j,k}^\hbr(x,y)\|\;  = \;
{\bf O}(\hbr^2)\; .$$
 Let us consider a function
 $\chi \; \in \; C^\infty (\bR )\ $ such that

 $\ \chi (t)=1\ $ if $\ |t|\leq 1/2\ $ and

 $\ \chi (t)=0\ $ if $\ |t|\ >1\; .$

 Then
 $\ (D_y^2+g(y))(1-\chi (|y|))\varphi_j(y)\; \in \; S(\bR ^m)\; ,$\\
 and Proposition \ref{thHorm} shows that
 $\  (1-\chi (|y|))\varphi_j(y)\; \in \;
 S(\bR ^m)\; . $\\
 As $\ D_y^2\varphi_j(y)=(\mu_j-g(y))\varphi_j(y)\ ,$ we get that
 \begin{equation}\label{phiP1}
 \forall \; k \; \in \; \bN \; ,\quad
 (1+|y|)^k[|\varphi_j(y)|^2+|D_y\varphi_j(y)|^2+|D_y^2\varphi_j(y)|^2]\; \in \;
 L^1(\bR ^m)\; .
 \end{equation}

The quantity $(\ccH\ -\ \tH_1)( u_{j,k}^\hbr(x,y))$ is,
 by  (\ref{deftH2}) , composed of 3 parts.
 According to  Lemma \ref{ppty} and the estimate (\ref{phiP1}),
 the two last parts are bounded in $L^2$-norm by
 $\ \hbr ^2C\; ,  \ (\mu_j\leq C\; )\; .$

To obtain a bound for the first part, we integrate by parts to get that
$$\| \frac{\nabla f(x)}{f(x)}D_x\psi^{\hbr}_{j,k}\| ^2
\; \leq \; C \left [ \| D_x^2\psi^{\hbr}_{j,k}\| \times \|
\frac{|\nabla f(x)|^2}{f^2(x)} \psi^{\hbr}_{j,k}\| \; +\; \|
D_x\psi^{\hbr}_{j,k}\| \times \| \frac{|\nabla f(x)|}{f(x)}
\psi^{\hbr}_{j,k}\| \right ]\; ,$$ and then we use again Lemma
\ref{ppty}.  Thus~: $\displaystyle \ \| \frac{\nabla
f(x)}{f(x)}D_x\psi^{\hbr}_{j,k}\|
\; \leq \; C \; .$ \\
According to estimate (\ref{phiP1}) we have finally
$\displaystyle \ \| \frac{\nabla
f(x)}{f(x)}D_x(yD_y)u^{\hbr}_{j,k}\| \leq  C  .$

 \section{ Middle energies}

We are going to refine the preceding results when $a\geq 2$ and $f(\infty)=\infty$.
It is possible then to get sharp localization near the   $\mu_j$'s for much higher
  values of $j$'s. More precisely we prove~:
 \begin{theorem}\label{thGround3} .
 We assume (\ref{Hypf}) with $\ f(\infty) \; =\;  \infty\; ,$
  (\ref{Hypg}) with $\ a\; \geq \; 2\ $ and
 with $\ g\; \in \; C^\infty (\bR ^m)\; .$

Let us consider $\ j\ $  such that
$\ \mu_j\; \leq \; \hbr ^{-2}\; ;$ \\
then for any integer $\ N\; ,$ there exists a constant $\ C $
depending only on $\ N\ $ such that, for any
$\ k\; \leq \; N\; ,$ there exists
 an eigenvalue
$\; \lambda_{jk}\; \in \; sp_d \; (\ccH )\ $ verifying
\begin{equation}\label{equaGround2}
|\; \lambda_{jk}\; -\; \lambda_k \left ( \hbr^2 D^2_x+\mu_jf^{2/(2+a)}(x)\right )\; |\; \leq \; C\mu_j \hbr ^{2}\; .
\end{equation}
Consequently, when $k=1\; ,$ we have
\begin{equation}\label{equaGround2b}
|\; \lambda_{j1}\; -\; \left [\mu_j\; +\; \hbr (\mu_j)^{1/2}
\frac{tr((\partial ^2f(0))^{1/2})}{(2+a)^{1/2}} \right ]\; |\; \leq \; C\mu_j \hbr ^{2}\; .
\end{equation}

\end{theorem}

{\bf Proof }~:

Let us define the class of symbols
$\ S(p^s(y,\eta ))\; ,\ s\; \in \; \bR \; , $ with
$\ p(y,\eta )\; =\; |\eta |^2+g(y)+1\; .$
$$q(y,\eta )\; \in \; S(p^s(y,\eta ))\quad
{\mathrm {iff}}\quad q(y,\eta )\; \in \; C^\infty (\bR ^m\times
\bR ^m)
$$
and for any $ \; \alpha \ $ and $\ \beta \ \in \; \bN ^m\; ,$
$$p^{-s}(y,\eta ) (|\eta |+1)^{-|\alpha |} (|y|+1)^{-|\beta |}
D_{\eta}^{\alpha}D_{y}^{\beta}q(y,\eta )\; \in \;
L^\infty (\bR ^{2m})\; .$$
For such a symbol
$\ q(y,\eta )\; \in \; S(p^s(y,\eta )\; ,$
 we define the operator $\ Q\ $
 on
$\ S(\bR ^m)\; :$
$$Qf(y)\; =\; (2\pi )^{-m}\int_{\bR ^{2m}}
q(\frac{y+z}{2},\eta )
e^{i(y-z)\eta}f(z)dz d\eta \; .$$

We will say that $\ Q\; \in \; OPS(p^s(y,\eta ))\; .$

It is  well known, (see \cite{Hor}) that
$\ (D_y^2+g(y))^s\; \in \; OPS(p^s(y,\eta ))\; .$

As $\ a\; \geq \; 2\; $, we get that $
\ yD_y\; \in \; OPS(p(y,\eta ))\; ,$ and then that
$\ yD_y(D_y^2+g(y))^{-1}\; \in \; OPS(1)\; $.

Therefore $\ yD_y(D_y^2+g(y))^{-1}\ $
and $\ (yD_y)^2(D_y^2+g(y))^{-2}\ $
are bounded operator on
$\ L^2(\bR ^m)\; ,$ and we get as a consequence the following bound~:
\begin{equation}\label{phiP1b}
\mu_{j}^{-1} \| yD_y\varphi_j\| \; +\;
\mu_{j}^{-2} \| (yD_y)^2\varphi_j\| \; \leq \; C\; .
\end{equation}
As in the proof of Theorem \ref{thGround2}, using
(\ref{phiP1b}) instead of (\ref{phiP1}), we get easily that
$$\| (\ccH  -\tH  )u^{\hbr}_{j,k}\|
 \; \leq \; C[\; \hbr ^2 \mu_j\; +\; \hbr ^3\mu_{j}^{3/2} ]
 \; \leq \; C\hbr ^2\mu_j\; ,$$
and then Theorem \ref{thGround3} follows.

\section{An application}

We consider a Schr\"odinger operator on $\ L^2(\bR _{z}^{d})\ $
with $\ d\geq 2\; ,$
\begin{equation}\label{defPh}
P^h \; =\; - h ^2 \Delta \; +\; V(z) \;
\end{equation}
with a real and regular potential $\ V(z)\ $ satisfying
\begin{equation}\label{hypV1}
\begin{array}{c}
V\; \in \; C^\infty (\bR ^d \; ;\ [0, +\infty [)\\
\liminf _{|z| \to \infty} \; V(z)\; >\; 0\\
\Gamma \; =\; V^{-1}(\{ 0\} )\quad {\rm is \ a\ regular\
hypersurface.}
\end{array}
\end{equation}
By hypersurface, we mean a submanifold of codimension $\ 1\; .$
Moreover we assume that $\ \Gamma \ $ is connected and that there exist
$\; m\; \in \; \bN ^\star\ \ {\rm and}\ \ C_0\; >\; 0$ such that
for any $ z$ verifying $ d(z,\ \Gamma )\; <\;
C_{0}^{-1}$ 
\begin{equation}\label{hypV2}
C_{0}^{-1}d^{2m}(z,\ \Gamma ) \; \leq \; V(z)\; \leq \; C_0
\ d^{2m}(z,\ \Gamma )
\end{equation}
$(\ d(E,F)\ $ denotes the euclidian distance between $\ E\ $ and
$\ F\; )\; .$

We choose an orientation on $\ \Gamma \ $ and  a  unit normal
vector $\ N(s)\ $\\
 on each $\ s\; \in \; \Gamma \; ,$
and then,  we can define the function on $\ \Gamma \; ,$
\begin{equation}\label{defF}
f(s)\; =\; \frac{1}{(2m)!}\left ( N(s)\frac{\partial}{\partial
s}\right )^{2m}V(s)\; , \quad \forall \; s\; \in \; \Gamma \; .
\end{equation}
Then by (\ref{hypV1}) and (\ref{hypV2}), $\ f(s)\; >\; 0\; ,\quad
\forall \; s\; \in \; \Gamma \; .$

Finally we assume that the function $\ f\ $ achieves its minimum on
$\ \Gamma \ $ on a finite number of discrete points:
\begin{equation}\label{hypVf1}
\Sigma_0\; =\; f^{-1}(\{ \eta_0\} )\; =\; \{ s_1,\ldots ,\;
s_{\ell_0}\} \; , \quad if \quad \eta_0\; =\; \min_{s\in
\Gamma}\; f(s)\; ,
\end{equation}
and the hessian of $\ f\ $ at each point $\ s_j\; \in \; \Sigma_0\
$ is non degenerated:

$\ \exists \; \eta_1\; >\; 0\ \ s.t.$
\begin{equation}\label{hypV2}
\frac 12 \langle d\left (\langle df\; ;\; w\rangle \right )\; ;\;
w\rangle (s_j)\; \geq \; \eta_1 |w(s_j)|^2\; , \quad \forall\; w\;
\in \; T\Gamma \; ,\ \forall \; s_j\; \in \; \Sigma_0\; .
\end{equation}\
If $\ g\; =\; (g_{ij})\ $ is the riemannian metric on
$\ \Gamma \ ,$ then $\ |w(s)|\; =\; (g(w(s),w(s)))^{1/2}\; . $
The hessian of $\ f\ $ at each $\ s_j\; \in \; \Sigma_0\; ,$ is
  the symmetric operator on $\ T_{s_j}\Gamma\; ,\ \
Hess(f)_{s_j}\; , $ associated to the two-bilinear form  defined
on
$\ T_{s_j}\Gamma\ $ by~:
\begin{equation}\label{defHess}
(v,w)\; \in \; (T_{s_j}\Gamma )^2\ \to \ \frac 12 \langle d\left (\langle
df\; ;\; {\widetilde v}\rangle \right )\; ;\; {\widetilde
w}\rangle (s_j) \; ,
\end{equation}
$\displaystyle \forall \; ({\widetilde v},{\widetilde w} )\; \in
(T\Gamma )^2\ \ s.t.\ \ \ ({\widetilde v}(s_j),{\widetilde
w}(s_j))\; =\; (v,w)\; . $

$\ Hess(f)_{s_j}\ $ has $\ d-1\ $ non negative eigenvalues
$$  \rho_1^2(s_j)\leq \; \ldots
\; \leq \; \rho_{d-1}^{2}(s_j)\; ,\quad \quad (\; \rho_j(s_j)\;
>\; 0)\; .$$ In local coordinates, those eigenvalues are the ones
of the symmetric matrix
$$
\frac{1}{2}G^{1/2}(s_j)\left (\frac{\partial ^2}{\partial x_k
\partial x_\ell }f(s_j) \right )_{1\leq k, \ell \leq d-1}
G^{1/2}(s_j)\; , \quad (\; G(x)\; =\; \left ( g_{k,\ell }(x)
\right )_{1\leq k, \ell \leq d-1}\; )\; .
$$
 The eigenvalues
$\ \rho_k^2(s_j)\ $ do not depend on the choice of coordinates. We
denote
\begin{equation}\label{defTrHess}
Tr^+ (Hess(f(s_j)))\; =\; \sum_{\ell =1}^{d-1}\rho_\ell (s_j)\; .
\end{equation}

We denote by $\ (\mu_j)_{j\geq 1}\ $ the increasing sequence of the
eigenvalues of the operator $\Di \ -\; \frac{d^2}{dt^2}\; +\;
t^{2m}\ $
on $\ L^2(\bR )\; ,$\\
and by $\ (\varphi _j(t)\; )_{j\geq 1}\ $ the associated orthonormal
Hilbert base of eigenfunctions.

\begin{theorem}\label{locTh1}
Under the above assumptions, for any $\ N\; \in \; \bN ^\star \;
,$ there exist $\ h_0\; \in \; ]0,1]\ $ and $\ C_0\; >\; 0\ $
such that,
if $\ \mu_j\; <<\; h ^{-4m/(m+1)(2m+3)}\; ,$ \\
and if $\ \alpha \; \in \; \bN ^{d-1}\ $ and $\ |\alpha |\; \leq
\; N\; ,$\\
 then $\ \forall \; s_\ell \;  \in \; \Sigma_0\;
,\quad \exists \; \lambda_{j\ell\alpha }^{h}\; \in \; sp_d(P^h )\ \
\  s.t.$
$$ \left | \;  \lambda_{j\ell\alpha }^{h}\; -\; h ^{2m/(m+1)}
   \left [ \eta^{1/(m+1)}_{0} \mu_j\; +\; h ^{1/(m+1)}\mu_{j}^{1/2}
 \; \cA_\ell (\alpha) \right
] \; \right |$$
$$ \; \leq \; h^2\mu_{j}^{2+3/2m}C_0\; ;$$
with $\displaystyle  \cA_\ell (\alpha) \; =\;
\frac{1}{\eta^{m/(2m+2)}_{0}(m+1)^{1/2}}
\left [ 2\alpha \rho(s_\ell ) \; +\; Tr^+ (Hess(f(s_\ell )))\right ] \; .$
\\
$(\alpha \rho (s_\ell )\; =\; \alpha_1\rho_1(s_\ell )+\ldots
\alpha_{d-1} \rho_{d-1}(s_\ell )\; )\; .$
\end{theorem}

\noindent
{\bf Proof~:}

 Let $\ \cO_0\; \subset \; \bR ^d\  $ be an open neighbourhood of
$\ s_l\; \in \; \Sigma_0\; $,
such that there exists $\ \phi\; \in \; C^\infty (\cO_0\; ;\ \bR)\ $
satisfying
\begin{equation}\label{defPhi}
\begin{array}{c}
\Gamma_0\; =\; \Gamma \; \cap \; \cO_0 \; =\;
\{ z\in \cO_0\; ;\ \phi (z)=0\} \; ;\\
| \nabla \phi (z) |\; =\; 1\; ,\quad
\forall \; z\; \in \; \cO_0\; .
\end{array}
\end{equation}
 After changing $\ \cO_0 \ $ into a smaller neighbourhood if necessary, we can find
$\ \tau \; \in \; C^\infty (\cO_0\; ;\ \bR ^{d-1})\ $ such that
 $\ \tau (s_l)\; =\; 0\ $ and $\ \forall \; z\; \in \; \cO_0\; ,$
\begin{equation}\label{defTauj}
\begin{array}{c}
\nabla \tau_j(z) . \nabla \phi (z)\; =\; 0\; ,\quad
\forall j=1,\ldots , d-1\\
\MR{rank}\{ \nabla \tau_1(z),\ldots ,\nabla \tau_{d-1}(z)\} \; =\;
d-1\; .
\end{array}
\end{equation}
Then $\ (x, y)\; =\; (x_1,\ldots ,x_{d-1},y)\; =\; (\tau_1,\ldots
,\tau_{d-1} ,\phi )\ $
are local coordinates in $\ \cO_0\ $ such that
\begin{equation}\label{Laplace}
\begin{array}{c}
\Delta \; =\;  |\tG |^{-1/2}\sum_{1\leq i, j \leq d-1}
\partial_{x_i}\left ( |\tG |^{1/2}\; \tG^{ij}\partial_{x_j}
\right )\; +\; |\tG |^{-1/2}\partial_y\left ( |\tG |^{1/2} \partial_y \right )\\
V\; =\; y^{2m}{\widetilde f}(x,y)\quad
\MR{with}\quad {\widetilde f}\; \in \; C^\infty (\cV_0)\; ;
\end{array}
\end{equation}
$\cV_0 \ $ is an open neighbourhood of zero in
 $\ \bR ^d\; ,$

$\displaystyle   \tG^{ij}(x,y)=\tG^{ji}(x,y)
\; \in \; C^\infty (\cV_0 ;\ \bR )\; ,\quad
|\tG |^{-1}\; =\; det\left ( \tG^{ij}(x,y) \right )\; >\; 0\; .$

$\ x\; =\; (x_1,\ldots ,x_{d-1})\ $ are local coordinates on
$\ \Gamma_0$\\
and the metric $\ g\; =\; (g_{ij})\ $ on $\ \Gamma_0\ $ is given by
$$
 \left ( g_{ij}(x)\right )_{1\leq i, j\leq d-1} \; =\; G(x)\; ,
 \quad \MR{with}\quad
(G(x))^{-1}\; =\; \left ( \tG^{ij}(x,0)\right )_{1\leq i, j\leq d-1}\; .$$
If $\ w\; \in \; C^2_0(\cO_0)\ $ then
\begin{equation}\label{newP}
\begin{array}{c}
P^h w\; =\; {\widehat P}^h u\quad \MR{with}\\
u\; =\; |\tG |^{1/4}w\quad \MR{and}\\
 {\widehat P}^h  \; =\; - h ^2\sum_{1\leq i, j\leq d-1}
\partial_{x_i} \left ( \tG^{ij}\partial_{x_j}
\right )\; - \; h ^2 \partial^{2}_{y} \; +\; V\; +\; h ^2V_0\; ,
\end{array}
\end{equation}
for some $\ V_0\; \in \; C^\infty (\cV_0\; ;\ \bR )\; .$

Let us write
\begin{equation}\label{newV}
V(x,y)\; =\; y^{2m}f(x)\; +\; y^{2m+1}f_1(x)\; +\;
y^{2m+2}{\widetilde f}_2(x,y)\; :
\end{equation}
$f(x)\; =\; {\widetilde f}(x,0)\ $ and $\ {\widetilde f}_2\; \in \;
C^\infty (\cV_0)\; .$

We perform the change of variable (\ref{changeV})  and the related
unitary transformation,
$$(x,y)\; \to \; (x,t)\; =\; (x, f^{1/(2(m+1))}(x)y)\; ,\quad
u\; \to \; v\; =\; f^{-1/(4(m+1)}u\; ,$$
 to get that
\begin{equation}\label{defHt}
\begin{array}{c}
 {\widehat P}^h  u\; =\; {\widehat Q}^h   v\quad \quad  \MR{with}\\
 {\widehat Q}^h \; =\; Q_0^h\; +\; t^{2m+1}f^0_1(x)\; +\; h ^2R_0\; +\;
+h ^2tR_1\; +\;
  t^{2m+2}{\widetilde f}^0_2\; :
\\
Q_0^h\; =\; -h^2
\sum_{1\leq i, j\leq d-1}\partial_{x_i}\left ( g^{ij}
\partial_{x_j} \right ) \; +\; f^{1/(m+1)}(x)
\left ( -h ^2\partial ^{2}_{t}\; +\; t^{2m} \right )
 \end{array}
 \end{equation}
 and
 $\displaystyle \quad R_0 = ta(x,t)(\partial_xf(x)\partial_x)\partial_t
  + b(x,t)t\partial_t +$
$$
 \sum_{ij}b_{ij}(x,t)\partial_{x_i}f(x)\partial_{x_j}f(x)(t\partial_t)^2
  + c(x,t)\; ,$$
 $\displaystyle R_1\; =\; \sum_{1\leq i,j\leq d-1}
 \partial_{x_i}\left ( \alpha_{ij}(x,t)\partial_{x_j}\right )\; ,$
 all coefficients are regular in a neighbourhood of the zero in
 $\ \bR ^d\; . $

Let $\ \mu_j\ $ be as in the theorem \ref{locTh1}.
We define
$\ h_j\; =\; h^{1/(m+1)}/\mu^{1/2}_j\; .$\\
Let $\ \cO_0^\prime \ $ be a bounded open neighbourhood of
zero in $\ \bR ^{d-1}\ $ such that
$\ {\overline \cO}_0^\prime \; \subset \cO_0 \cap \{ (x,0)\; ;\ x\in \bR ^{d-1}\}
\; .$ \\
We consider the Dirichlet operator
on $\ L^2(\cO_0^\prime )\; ,\ H^{h_j}_{0}\; :$
\begin{equation}\label{defHzero}
H^{h_j}_{0}\; =\;
-\; h_j^2 \;
\sum_{1\leq k, \ell \leq d-1} \partial_{x_k}\left ( g^{k\ell}(x)
\partial_{x_\ell} \right )\; +\; f^{1/(m+1)}(x)\; .
\end{equation}
It is well known, (see for example \cite{He1} or \cite{HeSj1},
that for any
$\ \alpha \; \in \; \bN ^{d-1}\ $ satisfying the assumptions
of the theorem \ref{locTh1},  one has:
$$\exists \; \lambda^{h}_{j,\alpha}\; \in \; sp\; (H^{h_j}_{0} )\quad
\MR{s.t.} \quad | \lambda^{h}_{j,\alpha} \; - \;
[ \eta_{0}^{1/(m+1)} \; +\; h_j \cA_l(\alpha ) |\; \leq \; h_{j}^{2}C\; ;
$$
$\cA_l(\alpha)\ $ is defined in theorem \ref{locTh1} in relation
with our $\ s_l\; \in \; \Sigma_0\; .$\\
$C\ $ is  a constant  depending only on $\ N\; .$
We will denote by $\ \psi^{h_j}_{j,\alpha}(x)\ $ any
associated eigenfunction with  a $\ L^2$-norm equal to  $1$ .
Let $\ \chi_0\; \in \; C^\infty (\bR)\ $ such that
$$ \chi_0(t)=1\quad \MR{if}\quad   |t|\leq 1/2\quad   \MR{and}\quad
 \chi (t)=0\quad  \MR{if}\quad
 | t|\geq 1\; .$$
We define the following function~:
 $$ \ u^{h}_{j,\alpha}(x,t)\;
=\; h^{-1/(2m+2)}\chi_0(t/\epsilon_0)
\psi^{h_j}_{j,\alpha}(x)\left [
\varphi_j(h^{-1/(m+1)}t)\; -\;
h^{1/(m+1)}F_j^h,(x,t)\right ]\; ,$$ 
with $$ F_j^h(x,t)= f^0_1(x)f^{-1/(m+1)}(x)\phi_j(h^{-1/(m+1)}t) ,$$ where $\displaystyle \phi_j\; \in \; S(\bR)\; $ is solution of~:\\
$-\frac{d^2}{dt^2} \phi_j(t)\; +\; (t^{2m}\; -\; \mu_j)\phi_j(t)\; =\; 
t^{2m+1}\varphi_j(t)\; ,$\\
and $\ \epsilon_0\; \in \; ]0,1]\; $ is a small enough constant,
but independent of $\ h \ $ and $\ j\; .$\\
$\phi_j\ $ exists because $\ \mu_j\ $ is a non-degenerated
 eigenvalue and the related eigenfunction $\varphi_j$ (see \ref{defmuk}) verifies 
 $\ \int_\bR t^{2m+1}\varphi_j^2(t)\; dt\; =\; 0$ , since it is a real even or odd  function.

Using the similar estimates as in chapter  $3\; ,$ one can get easily
that
$$
\mu_{j}^{-1} \| t\partial _t \varphi_j \| _{L^2(\bR )} \; +\;
\mu_{j}^{-2} \| (t\partial _t)^2 \varphi_j \| _{L^2(\bR )} \; \leq
\; C$$ and $\ \forall \; k\; \in \; \bN \; ,\quad \exists \; C_k\;
>\; 0\quad \MR{s.t.}\quad \mu_{j}^{-k/2m} \| t^k \varphi_j \|
_{L^2(\bR )} \; \leq \; C_k\; .$

It is well known that there exists $\ \epsilon_1\; >\; 0\ $ s.t.

$\ |\mu_j\; -\; \mu_\ell |\; \geq \; \epsilon_1\; , \quad \forall
\; \ell \; \neq \; j\; ,$ then the inverse of $\displaystyle \
-\frac{d^2}{dt^2}\; +\; t^{2m} \; -\; \mu_j\ $ is $L^2(\bR
)$-bounded by $\ 1/\epsilon_1\; ,$ (on the orthogonal of $\;
\varphi_j\; )\; .$
 So in the same way as in chapter $3\; ,$ we get also
that
$$
\mu_{j}^{-2-1/2m} \| t\partial _t \phi_j \| _{L^2(\bR )} \; +\;
\mu_{j}^{-3-1/2m} \| (t\partial _t)^2 \phi_j \| _{L^2(\bR )} \;
\leq \; C$$ and $\ \forall \; k\; \in \; \bN \; ,\quad \exists \;
C_k\; >\; 0\quad \MR{s.t.}\quad \mu_{j}^{-1-(k+1)/2m} \| t^k
\phi_j \| _{L^2(\bR )}
\; \leq \; C_k\; .$ \\
As in the proof of Theorem \ref{thGround3}, we get easily that
$$
\| [ {\widehat Q}^h\   -\; \mu_j \lambda^{h}_{j,\alpha} ]
\chi_0(|x|/\epsilon_0)u^{h_j}_{j,\alpha}(x,t)\| _{L^2(\cO_0)}
\; \leq\; h^2 \mu_{j}^{(4m+3)/2m}C$$ and \\
$\displaystyle |\; \| \chi_0(|x|/\epsilon_0)
u^{h_j}_{j,\alpha}(x,t)\| _{L^2(\cO_0)} \; -\; 1\; |\; =\; {\bf
O}(h^{1/(m+1)} \mu_{j}^{(2m+1)/2m} )\;
=\; \circ (1)\; .$\\
So the theorem \ref{locTh1} follows easily.

\begin{remark}\label{rmfinal}
If in Theorem \ref{locTh1} we assume that $\ j\ $ is also bounded
by $\ N\;  ,$ then, as in \cite{HeSj4}, we can get a full
asymptotic expansion
$$\lambda^{h}_{j\ell \alpha}\; \sim \; h^{2m/(m+1)}
\sum_{k=0}^{+\infty} c_{j\ell k \alpha}h^{k/(m+1)}\; ,$$
and for the related eigenfunction, a quasimode of the form
$$u^{h}_{j\ell \alpha}(x,t)\; \sim \;
c(h ) e^{ -\psi (x)/h^{1/(m+1)}}\chi_0(t/\epsilon_0)
\sum_{k=0}^{+\infty} h^{k/(2m+2)}a_{j\ell k\alpha}(x)\phi_{jk}(t/h^{1/(m+1)}) \; .$$
\end{remark}

\end{document}